# Evidence for a spinon Kondo effect in cobalt atoms on single-layer 1T-TaSe$_2$

Yi Chen[1,2,†], Wen-Yu He[3,†], Wei Ruan[1,2,4,†], Jinwoong Hwang[5,6,7], Shujie Tang[5,6,8,9,10], Ryan L. Lee[1], Meng Wu[1,2], Tiancong Zhu[1,2], Canxun Zhang[2,11], Hyejin Ryu[5,12], Feng Wang[1,2,11], Steven G. Louie[1,2], Zhi-Xun Shen[6,8], Sung-Kwan Mo[5], Patrick A. Lee[3], Michael F. Crommie[1,2,11,*]

[1]*Department of Physics, University of California, Berkeley, California 94720, USA*

[2]*Materials Sciences Division, Lawrence Berkeley National Laboratory, Berkeley, California 94720, USA*

[3]*Department of Physics, Massachusetts Institute of Technology, Cambridge MA 02139, USA*

[4]*Department of Physics, Fudan University, Shanghai 200438, China*

[5]*Advanced Light Source, Lawrence Berkeley National Laboratory, Berkeley, California 94720, USA*

[6]*Stanford Institute for Materials and Energy Sciences, SLAC National Accelerator Laboratory and Stanford University, Menlo Park, California 94025, USA*

[7]*Department of Physics, Pusan National University, Busan 46241, Korea*

[8]*Geballe Laboratory for Advanced Materials, Departments of Physics and Applied Physics, Stanford University, Stanford, California 94305, USA*

[9]*CAS Center for Excellence in Superconducting Electronics, Shanghai Institute of Microsystem and Information Technology, Chinese Academy of Sciences, Shanghai 200050, China*

[10]*School of Physical Science and Technology, Shanghai Tech University, Shanghai 200031, China*

[11]*Kavli Energy Nano Sciences Institute at the University of California Berkeley and the Lawrence Berkeley National Laboratory, Berkeley, California 94720, USA*

[12]*Center for Spintronics, Korea Institute of Science and Technology, Seoul 02792, Korea*

† These authors contributed equally to this work.

*e-mail: crommie@berkeley.edu

**Abstract:**

Quantum spin liquids (QSLs) are highly entangled, disordered magnetic states that arise in frustrated Mott insulators and host exotic fractional excitations such as spinons and chargons. Despite being charge insulators some QSLs are predicted to exhibit gapless itinerant spinons that yield metallic behavior in the spin channel. We have deposited isolated magnetic atoms onto single-layer (SL) 1T-TaSe$_2$, a gapless QSL candidate, to experimentally probe how itinerant spinons couple to impurity spin centers. Using scanning tunneling spectroscopy we observe the emergence of new, impurity-induced resonance peaks at the 1T-TaSe$_2$ Hubbard band edges when cobalt adatoms are positioned to have maximal spatial overlap with the Hubbard band charge distribution. These resonance peaks disappear when the spatial overlap is reduced or when the magnetic impurities are replaced with non-magnetic impurities. Theoretical simulations using a modified Anderson impurity model integrated with a gapless quantum spin liquid show that these resonance peaks are consistent with a Kondo resonance induced by spinons combined with spinon-chargon binding effects that arise due to QSL gauge-field fluctuations.

Gapless quantum spin liquids (QSLs) are predicted to act like "neutral metals" that exhibit metallic behavior in the spin channel despite being Mott insulators[1-9]. Evidence for this exotic metal-like behavior has been observed using thermal and magnetic measurements for several gapless QSL candidates[10-13]. These bulk probes, however, average over impurities and disorder, leading to inconsistent results for different samples[10,14-16] and/or alternative extrinsic explanations of the experimental signatures[17]. Local probe measurements, on the other hand, allow direct comparison of defect-free regions of a material with regions containing well-defined impurities[18,19]. Local magnetic moments are a particularly interesting type of impurity because their direct interactions with spinons allow them to potentially probe the effects of spinon itinerancy[20]. In a conventional metal such impurities are collectively screened by itinerant conduction electrons via the Kondo effect[21]. For a gapless QSL having a spinon Fermi surface, it has been predicted that itinerant spinons might also perform Kondo screening of magnetic impurities, yielding Kondo temperatures that scale similar to the conventional metallic case[20]. Some experimental evidence for spinon Kondo screening has been obtained from muon spin rotation studies of the QSL candidate Zn-brochantite ($ZnCu_3(OH)_6SO_4$) that show reduced magnetic moments arising from Cu-Zn intersite disorder[22]. Local measurement of individual Kondo impurities in a QSL, however, have not yet been reported.

Here we demonstrate the use of scanning tunneling microscopy/spectroscopy (STM/STS) to probe the Kondo response of individual magnetic impurities in a two-dimensional (2D) QSL. This type of measurement is complicated by the fact that QSLs are typically good insulators whereas STMs require nonzero electrical conductivity to function. We have overcome this challenge by using a new 2D gapless QSL candidate, single-layer (SL) 1T-$TaSe_2$, that is supported by a sheet of conductive graphene and so allows injected charge to be drained[23,24].

STM imaging allows us to identify defect-free patches of SL 1T-TaSe$_2$ where we have previously discovered long-wavelength spatial modulations indicative of a half-filled spinon Fermi surface[24]. In this work we are able to introduce individual magnetic impurities onto the clean 1T-TaSe$_2$ surface by vacuum-depositing a low coverage of Co atoms. By using STS at $T =$ 5 K we observe that Co atoms having maximal spatial overlap with the 1T-TaSe$_2$ $d$-orbital charge distribution exhibit new resonance peaks at *both* the upper and lower Hubbard band edges in the vicinity of Co impurities. Nonmagnetic Au atoms deposited onto the 1T-TaSe$_2$ surface do not show this unusual behavior. In order to understand why Co atoms on SL 1T-TaSe$_2$ generate new electron-hole bandedge resonances, we simulated our magnetic impurity system utilizing a modified Anderson model. Strong interaction effects were accounted for both in the substrate and in the impurity through a slave rotor mean-field theory approach employing a U(1) gapless QSL. Our calculations show that the experimentally observed resonance peaks are consistent with the formation of a Kondo screening cloud in the QSL spin channel that induces detectable impurity bound-state formation through emergent gauge-field fluctuations.

    The main experimental result reported here was obtained for individual magnetic Co atoms positioned at the center of the 1T-TaSe$_2$ charge density wave (CDW) unit cell (magnetism in Co adatoms for this configuration is supported by density functional theory (DFT) simulations (Methods)). As sketched in Fig. 1a, in the CDW phase of 1T-TaSe$_2$ the Ta atom lattice distorts (due to electron-phonon interactions[25]) to form a triangular superlattice of star-of-David (SOD) clusters, each composed of 13 Ta atoms. An STM image of a single Co atom sitting at the center of a SOD cluster can be seen in the STM topograph of Fig. 1b (obtained at a sample bias voltage of $V_b = -0.5$ V). At this bias voltage the SOD CDW superlattice can clearly be seen, and the Co adatom creates a bright protrusion ~1.4 Å high. Imaging the same region at a higher bias of $V_b =$

1.5 V reduces the CDW contrast, allowing the position of the Co atom to be more clearly determined (Fig. 1c).

An STS d$I$/d$V$ spectrum obtained for a single Co atom on 1T-TaSe$_2$ can be seen in Fig. 1d. Here the blue curve shows the spectrum obtained when the STM tip is held over the center of a pristine SOD cluster with no Co adatom (blue dot in Figs. 1b, c). Two peaks are seen that mark the lower Hubbard band (LHB) at $V_b$ = -0.34 V and the upper Hubbard band (UHB$_2$) at $V_b$ = 0.62 V for pristine SL 1T-TaSe$_2$, in agreement with previous results[23]. The upper Hubbard band is marked "UHB$_2$" because another Hubbard band feature (UHB$_1$) occurs at lower energy ($V_b$ = 0.17 V) but has little weight (i.e., local density of states (LDOS)) at the SOD center (Fig. 2b and Ref. [23]). The red curve shows the d$I$/d$V$ spectrum measured while placing the STM tip above the Co atom shown in Figs. 1b, c (red dot). Two outer peaks are seen that correspond to the LHB and UHB$_2$ features of the pristine surface. Strikingly, however, two new peaks are observed at $V_b$ = -0.15 V (labeled P$_1$, just above the LHB bandedge) and at $V_b$ = 0.53 V (labeled P$_2$, just below the UHB$_2$ bandedge). These two peaks are the main experimental finding of this paper. The P$_1$ and P$_2$ peaks are relatively narrow with a full-width at half max (FWHM) of 106 meV and 90 meV, respectively.

The Co electronic structure is sensitive to where on the surface the adatom sits, as shown in Fig. 2. Here local electronic structure variation can be seen in both our pristine surface spectroscopy and our "on-Co" spectroscopy, depending on atomic position. In general there are three types of positions that Co atoms can occupy (see sketch in Fig. 2a): (i) on-center, (ii) off-center, and (iii) midway (these positions are referenced to the center of a SOD unit cell). Co atoms occupying these three different positions can be seen in the topographs of Figs. 2c, d. One of the main differences between the three different positions is the strength of the pristine SL 1T-TaSe$_2$ Hubbard band LDOS, which is sketched in Fig. 2b. As shown previously[23], both the LHB and UHB$_2$ features exhibit high LDOS at the on-center position and decrease monotonically as one moves outward. The UHB$_1$ feature, in contrast, peaks at locations away from the SOD center. This behavior is reflected in the pristine (i.e., off-Co) d$I$/d$V$ spectra which show the UHB$_1$ feature absent at the on-center position, but appearing more strongly for the off-center and midway positions (blue curves, Fig. 2e).

The on-Co d$I$/d$V$ spectra in Fig. 2e (red curves) differ significantly when the Co atom sits at different positions. While the P$_1$ and P$_2$ features appear strongly for the on-center position, they are completely absent when Co atoms occupy either the off-center or midway positions. At the off-center position Co atoms exhibit features that correspond to the LHB, UHB$_1$, and UHB$_2$ features, but the UHB$_1$ peak observed at $V_b = 0.25$ V is slightly offset from the pristine UHB$_1$ feature at $V_b = 0.17$ V. At the midway position Co atoms exhibit features corresponding to LHB and UHB$_2$, but show no features corresponding to UHB$_1$. All of the spectroscopic features shown in Figs. 2c-e were reversibly reproduced when we shifted individual Co atoms from one position to another for all three positions using atomic manipulation (see Supplementary Note 3). This helps to clarify that the spectroscopic features highlighted above arise from individual Co atoms and are not due to other types of surface defects.

As a control measure we performed these same types of measurements using a nonmagnetic impurity, Au. Au atoms were deposited onto SL 1T-TaSe$_2$ under the same general conditions that we used for depositing Co atoms. An STM topograph of a single Au atom on SL 1T-TaSe$_2$ can be seen in the inset to Fig. 3. Au atoms show a preference for binding at the center of SOD clusters. Occasionally Au atoms were found at other positions on the surface, but in those cases the atoms were unstable under STM spectroscopy conditions (the only stable position was the on-center position). The results of STM d$I$/d$V$ spectroscopy performed with the STM tip held over Au atoms in the on-center position can be seen in Fig. 3 (red curve). The Au adatom spectrum shows no new resonances compared to the reference spectrum (blue curve, using the same tip) and notably does not show either the $P_1$ or the $P_2$ peak that was observed for Co adatoms. Features in the Au adatom spectrum corresponding to the pristine LHB and $UHB_2$ resonances can be seen, although the LHB feature appears to be shifted to slightly lower voltage for the on-Au spectrum ($V_b$ = -0.46 V) compared to the pristine SL 1T-TaSe$_2$ spectrum ($V_b$ = -0.31 V).

The appearance of two new resonances symmetrically aligned adjacent to the Hubbard bandedges as seen for Co atoms in the on-center position, is an unusual type of defect behavior. For conventional semiconductors such bandedge resonances do occur for shallow donor and acceptor impurities, but typically at only one bandedge. Donor states reside below the conduction bandedge and acceptor states reside above the valence bandedge, but it is abnormal for a single type of defect to simultaneously exhibit a resonance at both energies. Defect behavior in Mott insulators is more poorly understood than in conventional semiconductors, but the behavior we observe is unusual even for Mott insulators. In "conventional" Mott insulators such as cuprates and iridates, for example, impurities have generally been found to induce broad in-gap states that fill up the Mott-Hubbard gap and metallize the system through a doping mechanism[26-31]. In bulk 1T-TaS$_2$ (which has many similarities to SL 1T-TaSe$_2$), the Mott insulating state has been shown to partially or fully collapse upon the introduction of defects, K adatoms, or the substitution of S with Se[32-34]. The appearance of symmetrically-placed resonances near the Hubbard bandedges, however, has not previously been seen.

We are able to explain the observed behavior of Co adatoms at the surface of 1T-TaSe$_2$ by the appearance of the Kondo effect for a magnetic impurity in contact with a gapless QSL. For the "conventional" Fermi liquid Kondo effect, scattering between itinerant conduction electrons and a localized impurity spin leads to a manybody singlet state whose signature is a narrow "Kondo resonance" at $E_F$. Such features have been observed by STM for individual magnetic impurities on metal surfaces and exhibit a Fano lineshape in the electronic LDOS[18,35]. For a gapless QSL the Kondo behavior is predicted to be strikingly similar to the Fermi liquid case if one considers only the spin channel[20] (since spinons in a gapless QSL behave like the fermionic inhabitants of a "neutral metal"). The problem lies in how to experimentally detect a Kondo resonance that exists only in the spin channel of a Mott insulator. A solution is to take into account both the spin and charge channels simultaneously. Since conventional STMs can only tunnel *electrons* into a material, these electrons must decompose into both spinons and chargons once they are inside a QSL (since these are the elementary excitations of QSLs[3]). STM electron tunnel current into a QSL can thus be expressed as a convolution of the spinon and chargon density of states[9,24,36,37]. If one wants to use an STM to detect a Kondo resonance in the spin channel of a QSL, one must therefore consider how it will appear when convoluted with the simultaneously occurring chargon density of states.

Such reasoning led us to calculate *both* the spinon *and* chargon density of states for a magnetic impurity in contact with a QSL. To accomplish this we utilized a modified Anderson model incorporating a U(1) QSL exhibiting a spinon Fermi surface (the model parameters are mostly determined by experimental spectra, see Methods, Supplementary Note 1, and Ref. [38]). The model was solved using a slave-rotor mean-field theory where electrons are decomposed into spin-1/2 auxiliary fermions (the spin channel) and charged spinless bosons (the charge channel)[8,37,39]. The auxiliary fermions and charged bosons within the QSL (not including the impurity) are referred to as spinons and chargons, respectively. The solution of the Anderson model in the spin channel yields a Kondo resonance peak in the density of states (DOS) of the impurity auxiliary fermion at the Fermi level of the QSL spinon band, as shown in Fig. 4a. In the QSL spinon spectrum the Kondo resonance manifests as a dip at the Fermi level (Fig. 4b), which can be rationalized as level repulsion in a standard Fano-type scenario (i.e., involving a discrete state in resonance with a continuum)[35,40].

In the charge channel the impurity charged boson hybridizes with QSL chargon states, resulting in an intermixing of their DOS spectra (Figs. 4c, d). For example, in addition to having the usual bare impurity levels ("$E_d$" and "$E_d + U$"), the impurity exhibits new spectral features that appear at the QSL Hubbard band energies (Fig. 4c, dashed lines) which are inherited from the QSL chargon spectrum. Similarly, the QSL chargon DOS spectrum exhibits weak peaks at $E_d$ and $E_d + U$ (Fig. 4d, dashed lines) that originate from the bare impurity levels.

In order to obtain the final predicted tunneling DOS for electrons, the spin and charge channels must be combined via convolution (see Methods and Refs. [9,24,36,37]) into a *total* electron channel. The result of this convolution can be seen in Fig. 4 for both the impurity (Fig. 4e) and the QSL continuum (Fig. 4f). The resulting impurity electronic DOS spectrum (Fig. 4e) exhibits a step function at the Hubbard band edges (i.e., at the energies $\pm\Delta_{Mott}$) that is more abrupt than the linear dispersion seen in the absence of impurity (Fig. S4c). However, there are no peaks at the bandedges that resemble the experimental $P_1$, $P_2$ features of Fig. 1d.

Explaining the experimentally observed side-peaks in the on-Co STS requires the incorporation of more realistic effects into our model, namely the inclusion of longitudinal gauge-field fluctuations (see Methods, Supplementary Note 2, and Ref. [38]). The reason for this is that in a U(1) QSL with a spinon fermi surface, both spinons and chargons couple via an emergent U(1) gauge field[6]. The gauge field produces an effective binding interaction $U_b$ [41,42] since spinons and chargons have opposite gauge charge. For a pristine QSL the inclusion of this spinon-chargon binding merely shifts the positions of the LHB and UHB peaks (Fig. S4). For a magnetic impurity embedded in a QSL, however, the gauge binding causes the spinon Kondo screening cloud to trap chargons and thereby form bound electronic states in the QSL (see sketch in Fig. 5a). Since the localized Kondo spinon screening cloud is attractive to chargons in both the holon and doublon branches, spinon-chargon bound states form at energies *both* above the holon branch and below the doublon branch (Fig. 5b), in analogy to acceptor and donor bound states in a semiconductor.

Inclusion of this effect into our model causes a pile-up of spectral weight near the band edges. The pile-up appears at a small binding interaction $U_b = 0.08$ eV and increases smoothly with increasing $U_b$, as shown in Fig. S5. The value $U_b = 0.19$ eV results in a pair of resonance peaks at the Hubbard band edges of the total impurity electronic DOS that best fits our data (see Fig. 5d and compare with Fig. 5c where no binding effects are present). The $P_1$, $P_2$ side-peaks in our experimental STM spectrum (Fig. 1d) can thus be understood to reflect spinon-chargon bound states induced by the spinon Kondo cloud around a magnetic impurity. We find that varying model parameters such as the Coulomb repulsion, *U,* and hybridization, *V,* modifies the spinon Kondo temperature ($T_K$) but does not qualitatively change the physical picture presented here. So long as the impurity is in the Kondo screening regime with $T < T_K$ then spin correlations will always induce bandedge resonance peaks at the site of the impurity when the gauge binding interaction, $U_b$, is sufficiently large. This mechanism also applies to impurities with spin higher than $s = ½$ that could potentially be underscreened[43] by the Kondo effect. While $U_b$ is the most essential parameter in this physical framework for determining how bandedge resonance peaks arise when $T < T_K$, it currently can only be determined by comparing theoretical simulations with experimental data. The value $U_b$ = 0.19 eV that we extract from our data in this way is strong enough to induce pronounced spinon-chargon bound states in the vicinity of the magnetic impurity, but is not strong enough to bind a spinon and chargon in the absence of an impurity (Fig. S4).

Our observed Co adatom position-dependence in the CDW unit cell is consistent with the orbital nature of the 1T-TaSe$_2$ Hubbard bands within this overall scenario. Because the SOD unit cell is composed of 13 Ta atoms (each containing one $d$-electron[23]), the fundamental building block of the SL 1T-TaSe$_2$ Mott bands is more akin to a molecular orbital than to an individual atomic state. This is manifest, for example, in the position-dependence of UHB$_1$ for the pristine material (which is stronger *away* from the SOD center) and in UHB$_2$ (which is stronger *near* the SOD center) (see Fig. 2b). Our calculations do not include this level of orbital complexity, but the experimental appearance of a strong Kondo effect involving UHB$_2$ for Co in the on-center position is qualitatively consistent with this reasoning. When a Co adatom is moved away from the on-center position we expect coupling between the impurity and the QSL charge distribution to reduce, thus reducing the Kondo temperature below our detection limits (Fig. S6). Similar reasoning could explain why impurity coupling to UHB$_1$ does not result in a detectable Kondo effect. The placement of nonmagnetic atoms (such as Au) onto the on-center position is similarly not expected to result in Kondo-induced side-peaks due to the lack of spin-based exchange coupling, consistent with the experimental data (Fig. 3).

In conclusion, we have deposited isolated magnetic adatoms onto the single-layer QSL candidate 1T-TaSe$_2$. Magnetic impurities having large hybridization with the underlying

Hubbard band LDOS develop a pair of resonance peaks at the Hubbard band edges, in agreement with the impurity electronic density of states obtained from a modified Anderson model incorporating a U(1) gapless QSL. This provides experimental evidence supporting the existence of a spinon-based Kondo effect in a gapless QSL. Our ability to perform atomically-precise impurity manipulation and spectroscopy with the tip of an STM might be applied to other layered QSL systems in order to gain microscopic control over other exotic excitations such as non-abelian anyons[44-47].

**Acknowledgments**

This research was supported by the U.S. Department of Energy, Office of Science, National Quantum Information Science Research Centers, Quantum Systems Accelerator (STM/STS measurements) and the Advanced Light Source (sample growth) funded by the Director, Office of Science, Office of Basic Energy Sciences, Materials Sciences and Engineering Division, of the US Department of Energy under Contract No. DE-AC02-05CH11231. Support was also provided by National Science Foundation award DMR-1807233 (topographic characterization). The work at the Stanford Institute for Materials and Energy Sciences and Stanford University (surface treatment) was supported by the DOE Office of Basic Energy Sciences, Division of Material Science. P.A.L. acknowledges support by DOE Basic Energy Science award number DE-FG02-03ER46076 (theoretical QSL analysis). H.R. acknowledges support from a National Research Foundation of Korea (NRF) grant funded by the Korea government (MSIT) (No. 2021R1A2C2014179) (growth characterization).

**Author contributions**

Y.C., W.R., P.A.L., and M.F.C. initiated and conceived this project. Y.C., W.R., R.L., T.Z., and C.Z. carried out STM/STS measurements under the supervision of M.F.C. J. H., S.T., and H.R. performed sample growth under the supervision of Z.-X.S. and S.-K.M. W.Y.H. performed slave rotor calculations and theoretical analysis under the supervision of P.A.L. M.W. performed DFT calculations under the supervision of S.G.L. Y.C., W.Y.H., W.R., and M.F.C. wrote the manuscript with the help from all authors. All authors contributed to the scientific discussion.

**Competing interests**

The authors declare no competing interests.

**Data availability**

All data that support the plots within this paper and other findings of this study are available from the corresponding author upon reasonable request.

**Code availability**

The codes used in this study are available from the corresponding author upon reasonable request.

**Methods**

**Sample growth**

Single-layer 1T-TaSe$_2$ films were grown on epitaxial bilayer graphene terminated 6H-SiC(0001) as well as cleaved HOPG substrates in a molecular beam epitaxy chamber operating at ultrahigh vacuum (UHV, base pressure $2\times10^{-10}$ Torr) at the HERS endstation of Beamline 10.0.1, Advanced Light Source, Lawrence Berkeley National Laboratory. High purity Ta (99.9%) and Se (99.999%) were evaporated from an electron-beam evaporator and a standard Knudsen cell, respectively, with a Ta:Se flux ratio set between 1:10 and 1:20 and a substrate temperature of 660 °C. The growth process was monitored by reflection high-energy electron diffraction. Before taking the films out of vacuum for STM/STS measurements, Se capping layers with ~10 nm thickness were deposited onto the samples for passivation. These were later removed by UHV annealing at ~180 °C for 1 hour.

**STM/STS measurements**

STM/STS measurements were performed using a commercial CreaTec STM/AFM system at $T = 5$ K under UHV conditions. To avoid tip artifacts, STM tips were calibrated on a Au(111) surface by measuring its herringbone surface reconstruction and Shockley surface state before all STM/STS measurements. STS d$I$/d$V$ spectra were obtained using standard lock-in techniques with a small bias modulation at 401 Hz. High purity Co wire (99.99%) was used to evaporate Co atoms from an electron-beam evaporator onto cold samples kept in the STM head (to avoid adatom clustering). Co evaporation was first calibrated on Au(111) substrates where the existence of isolated Co atoms was confirmed through their known Kondo resonance features[18,35]. Single-layer 1T-TaSe$_2$/BLG on SiC was then transferred into the STM and brought to thermal equilibrium before performing another Co evaporation using identical deposition parameters. High purity Au wire (99.99%) was used to evaporate Au atoms via direct heating onto cold samples kept in the STM head (i.e., by wrapping the Au wire around a W filament heater).

**Anderson impurity model on QSL**

The following Hamiltonian was used to describe an atomic impurity with onsite repulsion $U$ embedded in a QSL:

$$H = \sum_\sigma E_d d_\sigma^\dagger d_\sigma + U d_\uparrow^\dagger d_\uparrow d_\downarrow^\dagger d_\downarrow + V \sum_\sigma (c_{\sigma,R}^\dagger d_\sigma + d_\sigma^\dagger c_{\sigma,R}) + H_h \;.$$

Here $E_d$ is the onsite energy, $\sigma = \uparrow/\downarrow$ is the spin index, $V$ is the coupling between the impurity and the QSL electronic state at the nearest neighbor site $R$, and $H_h = t \sum_{r,r',\sigma} c_{r,\sigma}^\dagger c_{r',\sigma} + U_{QSL} \sum_r c_{r,\uparrow}^\dagger c_{r,\uparrow} c_{r,\downarrow}^\dagger c_{r,\downarrow}$ is the half-filled Hubbard model for the QSL.

The half-filled Hubbard model for the QSL has a slave rotor mean field solution, which fractionalizes the electrons in the QSL as $c_{\sigma,r}^\dagger = f_{\sigma,r}^\dagger X_r$, $c_{\sigma,r} = f_{\sigma,r} X_r^\dagger$, with $f_{\sigma,r}^{(\dagger)}$, $X_r^{(\dagger)}$ being the annihilation (creation) operators of the spinons and chargons, respectively. The spinon and chargon band dispersions in the QSL are taken to be

$$\xi_k = 2t_f \left(2\cos\tfrac{1}{2}k_x a \cos\tfrac{\sqrt{3}}{2}k_y a + \cos k_x a\right) - \mu_f,$$

$$\varepsilon_k = 2t_X \left(2\cos\tfrac{1}{2}k_x a \cos\tfrac{\sqrt{3}}{2}k_y a + \cos k_x a - 3\right) + \Delta_{Mott}.$$

Here $\Delta_{Mott} = 0.25$ eV is fixed by the experimentally observed Mott gap size (which coincides with $U_{QSL}/2$ in the atomic limit). The hopping parameters $t_f = 0.03$ eV and $t_X = -0.03$ eV are determined by the experimentally observed bandwidth of pristine SL 1T-TaSe$_2$, while the half-filling requirement of the spinon band fixes the chemical potential to be $\mu_f = -0.024$ eV.

In the slave rotor mean field formalism, the electron operators of the impurity are rewritten as $d_\sigma^\dagger = a_\sigma^\dagger X_d$, $d_\sigma = a_\sigma X_d^\dagger$. For the impurity electron, $a_\sigma^{(\dagger)}$ annihilates (creates) a spin-1/2 auxiliary fermion that represents the spin degree of freedom of the impurity electron and $X_d^{(\dagger)}$ annihilates (creates) a spinless charged boson that corresponds to its charge degree of freedom. The coupling between the Anderson impurity and the QSL was solved using the slave rotor mean field so that the Green's functions for the spin-1/2 auxiliary fermion and the charged boson at the impurity could be obtained. Here the bare Co Coulomb energy $U$ and the hopping $V$ between the Co atom and the QSL are taken to be $U = 5$ eV and $V = 0.23$ eV, which yield a Kondo temperature of $T_K$ = 19 K. Variation of $U$ and $V$ changes the Kondo temperature but does not significantly affect our main results (i.e., the appearance of bandedge resonance peaks in the Kondo regime with $T < T_K$). More details regarding the Green's function constructed from the slave rotor mean field theory can be found in Supplementary Note 1 and reference[38].

**Spectral function calculations**

The electronic spectral function (i.e., LDOS) at the impurity and the QSL is the imaginary part of the retarded Green's function. For the impurity electron the thermal Green's function is given by the correlation function

$$G_{d,\sigma}^0(i\omega_n) = -\langle d_\sigma(i\omega_n)d_\sigma^\dagger(i\omega_n)\rangle_0 = -\frac{1}{\beta}\sum_{\nu_n}\langle a_\sigma(i\omega_n+i\nu_n)a_\sigma^\dagger(i\omega_n+i\nu_n)\rangle_0 \langle X_d(i\nu_n)X_d^\dagger(i\nu_n)\rangle_0.$$

and similarly the thermal Green's function for the QSL electron at $\boldsymbol{R}$ is

$$G_{c,\sigma}^0(i\omega_n,\boldsymbol{R},\boldsymbol{R}) = -\langle c_\sigma(i\omega_n)c_\sigma^\dagger(i\omega_n)\rangle_0 = -\frac{1}{\beta}\sum_{\nu_n}\langle f_{\sigma,\boldsymbol{R}}(i\omega_n+i\nu_n)f_{\sigma,\boldsymbol{R}}^\dagger(i\omega_n+i\nu_n)\rangle_0 \langle X_{\boldsymbol{R}}(i\nu_n)X_{\boldsymbol{R}}^\dagger(i\nu_n)\rangle_0.$$

Here the $\omega_n$ and $\nu_n$ are the fermionic and bosonic Matsubara frequencies respectively, and $\langle\cdots\rangle_0$ means the thermal average calculated through the Anderson impurity Hamiltonian $H$. Analytic continuation $i\omega_n \to \omega + i0^+$ yields the corresponding retarded Green's function. The impurity electronic DOS shown in Fig. 4e as well as the QSL electronic DOS shown in Fig. 4f are given by $-\frac{1}{\pi}\mathrm{Im}G_{d,\sigma}^0(\omega)$ and $-\frac{1}{\pi}\mathrm{Im}G_{c,\sigma}^0(\omega)$ respectively. The detailed forms of these correlation functions are given in Supplementary Note 1.

Chargons and spinons in the QSL carry opposite gauge charge and so longitudinal gauge fluctuations yield an effective binding between them[22]:

$$U_b \sum_\sigma f_{\sigma,\boldsymbol{R}}^\dagger f_{\sigma,\boldsymbol{R}}\left(a_{\boldsymbol{R}}^\dagger a_{\boldsymbol{R}} - b_{\boldsymbol{R}}^\dagger b_{\boldsymbol{R}}\right),$$

with $a_R^{(\dagger)}, b_R^{(\dagger)}$ being the holon and doublon that compose the chargon $X_R^\dagger = a_R + b_R^\dagger, X_R = a_R^\dagger + b_R$. Incorporating the spinon-chargon binding interaction into the Anderson impurity Hamiltonian results in a modified correlation function that yields the final impurity electron thermal Green's function:

$$G_{d,\sigma}(i\omega_n) = -\langle d_\sigma(i\omega_n) d_\sigma^\dagger(i\omega_n)\rangle.$$

The impurity DOS spectrum is given by $-\frac{1}{\pi}\mathrm{Im} G_{d,\sigma}(i\omega_n \to \omega + i0^+)$ and exhibits a pair of resonance peaks at the Hubbard bandedges as shown in Fig. 5d. The detailed form of $G_{d,\sigma}(i\omega_n)$ can be found in Supplementary Note 2.

**DFT simulation of Co magnetic moment**

First-principles calculations of Co on the CDW phase of single-layer 1T-TaSe$_2$ were performed using density functional theory (DFT) as implemented in the Quantum ESPRESSO package. A slab model with 16 Å vacuum layer was adopted to avoid interactions between periodic images. We employed optimized norm-conserving Vanderbilt pseudopotentials (ONCVPSP) including Ta 5s and 5p semicore states (with a plane-wave energy cutoff of 90 Ry) as well as the Perdew-Burke-Ernzerhof (PBE) exchange-correlation functional in the generalized gradient approximation (GGA). To reduce the computational cost to a reasonable level, every CDW unit cell was assumed to host a Co atom above its central Ta atom, and the structure was fully relaxed at the DFT-PBE level until the force on each atom was less than 0.02 eV/Å. Onsite Hubbard interactions were added through the simplified rotationally invariant approach using $U = 2$ eV for each Ta atom and $U = 3$ eV for each Co atom[23,48,49]. DFT simulation results show that the Co atom on single-layer 1T-TaSe$_2$ slightly n-dopes the system and hosts a magnetic moment of 1.86 μB.

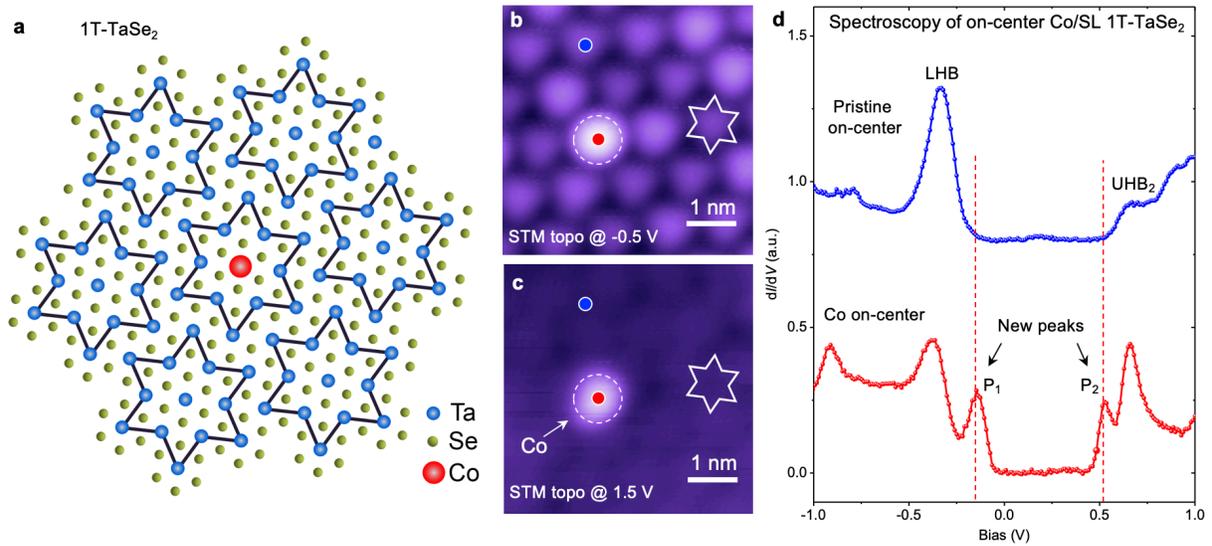

**Fig. 1. Co adatoms in the on-center position on single-layer 1T-TaSe₂ show new resonance peaks at Hubbard band edges. a**, Schematic shows 1T-TaSe₂ structure, including star-of-David CDW and on-center Co adatom position. **b**, STM topograph acquired at moderate negative sample bias voltage shows the Co adatom (white dashed line) as well as the star-of-David CDW superlattice (white solid line) for SL 1T-TaSe₂ ($V_b$ = -0.5 V, $I_t$ = 10 pA). **c**, STM topograph of same area as **b** acquired at high positive sample bias voltage shows only the Co adatom (white dashed circle) ($V_b$ = 1.5 V, $I_t$ = 10 pA). Comparison between **b** and **c** confirms that the Co adatom is "on-center" (i.e., located at the center of the underlying CDW cell). **d**, Blue curve shows a typical pristine SL 1T-TaSe₂ STM d$I$/d$V$ spectrum obtained at the on-center position of a CDW unit cell. Features corresponding to the Hubbard bandedges are marked LHB and UHB₂. Red curve shows a d$I$/d$V$ spectrum of a single Co adatom in the on-center position on SL 1T-TaSe₂ (red dot in **b** and **c**). Two new resonance peaks (labeled $P_1$, $P_2$) appear near the Hubbard band edges (dashed lines) for on-center Co ($V_b$ = -1 V, $I_t$ = 20 pA, $V_{r.m.s.}$ = 20 mV).

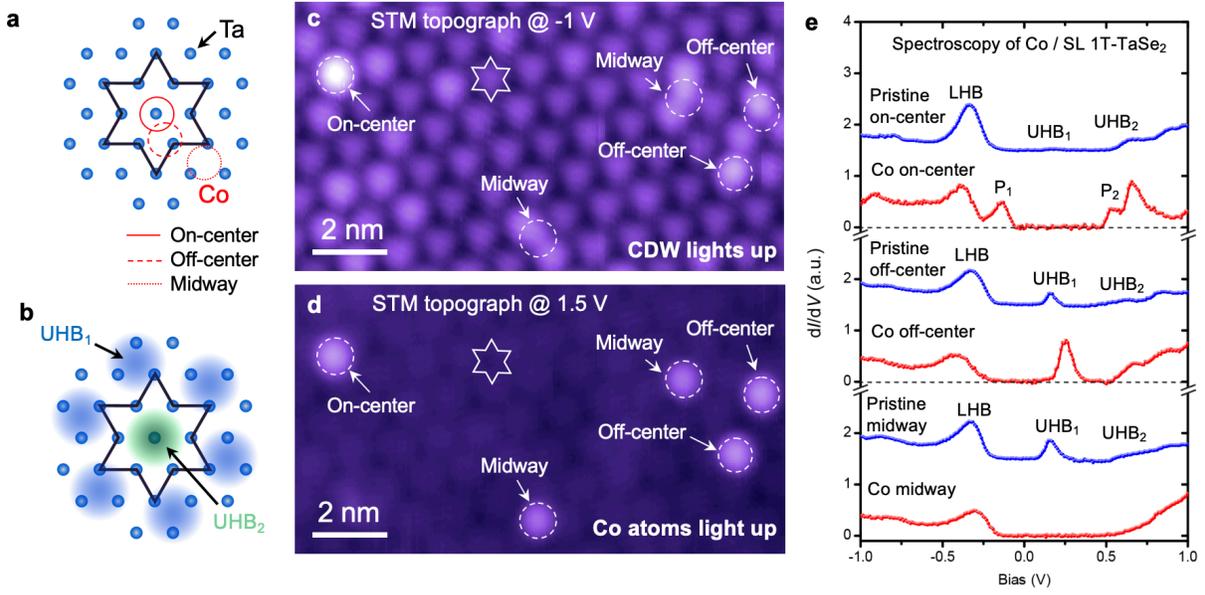

**Fig. 2. Electronic behavior of Co adatoms at different locations on SL 1T-TaSe$_2$. a**, Schematic shows three possible locations for Co atoms in the star-of-David CDW unit cell for SL 1T-TaSe$_2$. **b**, Sketch shows spatial distribution of LDOS in UHB$_1$ (blue) and UHB$_2$ (green) (from ref. [23]). The LHB LDOS coincides spatially with the UHB$_2$ LDOS. **c**, STM topograph taken at moderate negative sample bias shows Co adatoms (dashed circles) and SL 1T-TaSe$_2$ CDW lattice simultaneously ($I_t$ = 5 pA). **d**, STM topographic taken in same region as **c** at high positive sample bias shows Co adatom locations (dashed circles) without CDW background ($I_t$ = 5 pA). **e**, STM d$I$/d$V$ spectra of Co adatoms located at midway, off-center, and on-center positions compared to pristine SL 1T-TaSe$_2$ spectra taken at similar positions. New P$_1$ and P$_2$ resonance peaks at Hubbard band edges emerge only for Co adatoms at the on-center position ($V_b$ = -1 V, $I_t$ = 20 pA, $V_{r.m.s.}$ = 10 mV).

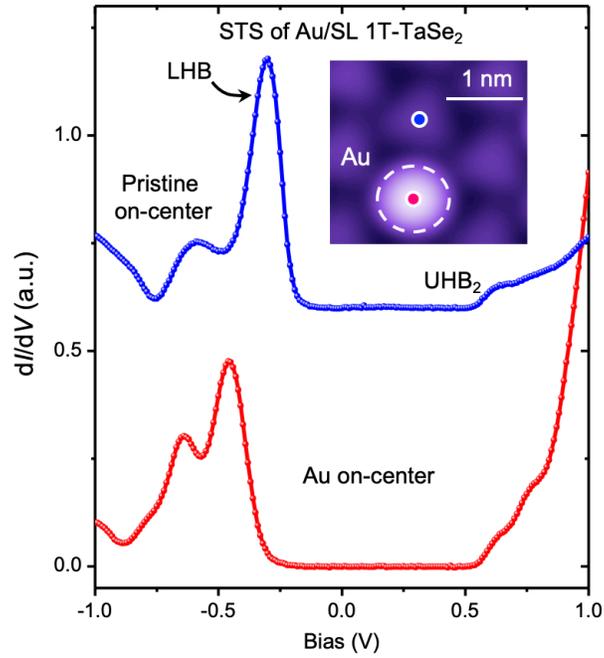

**Fig. 3. STM spectrum of a single non-magnetic impurity (Au) on SL 1T-TaSe$_2$.** STM d$I$/d$V$ spectrum of a single Au adatom at the on-center position on SL 1T-TaSe$_2$ (red curve) compared to a d$I$/d$V$ spectrum of pristine SL 1T-TaSe$_2$ taken at the same position in the CDW lattice with the same STM tip (blue curve) (spectra obtained at color-coded positions in the inset). No new resonance peaks at the Hubbard band edges are observed for on-center Au ($V_b$ = -1 V, $I_t$ = 50 pA, $V_{r.m.s.}$ = 20 mV). Inset: STM topograph shows an on-center Au adatom (white dashed circle) on SL 1T-TaSe$_2$ ($I_t$ = 5 pA, $V_b$ = -1 V).

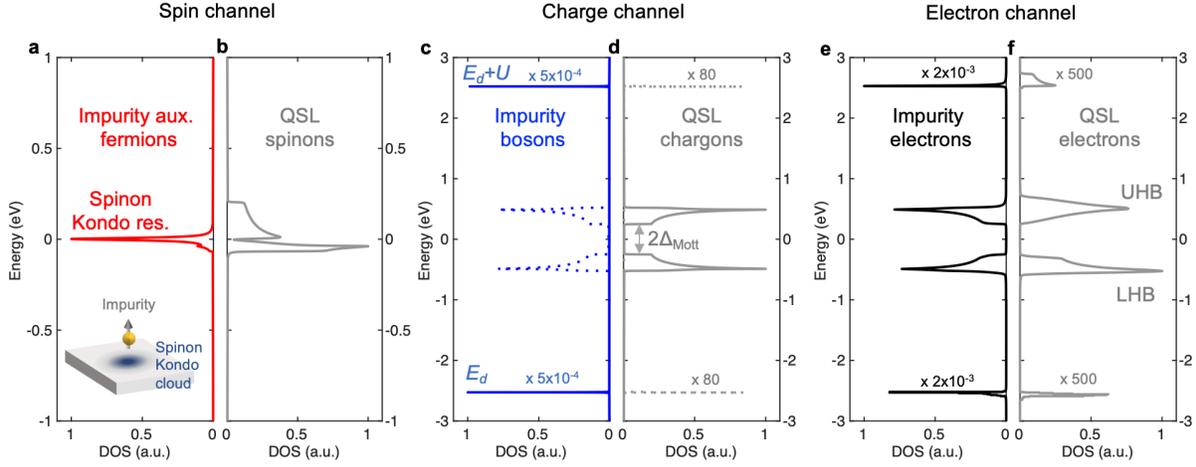

**Fig. 4. Calculated spinon Kondo effect.** The Kondo effect for a magnetic impurity in contact with a QSL was modeled using a modified Anderson model coupled to a U(1) QSL having a spinon Fermi surface. The model was solved using a slave-rotor mean-field technique whereby electrons are decomposed into auxiliary fermions and charged bosons (gauge fluctuations are ignored at this level). **a-b**, Resulting theoretical DOS spectra of impurity auxiliary fermions and QSL auxiliary fermions ("spinons"). Impurity auxiliary fermions show a sharp resonance at the Fermi level due to Kondo screening from itinerant QSL spinons. QSL spinon DOS shows a dip at $E_F$ due to level-repulsion from the Kondo resonance. Inset: sketch represents the spinon Kondo screening cloud induced by a magnetic impurity in a QSL. **c-d**, Theoretical DOS spectra of impurity charged bosons and QSL charged bosons ("chargons"). In addition to its intrinsic bare impurity levels ("$E_d$" and "$E_d + U$"), the impurity charged boson inherits Hubbard band features due to hybridization with QSL chargons (dashed curves). A similar effect occurs for the QSL chargons due to hybridization with the bare impurity levels. **e-f**, Theoretical *total* electron DOS spectra of impurity and QSL obtained by convolving spin and charge channels.

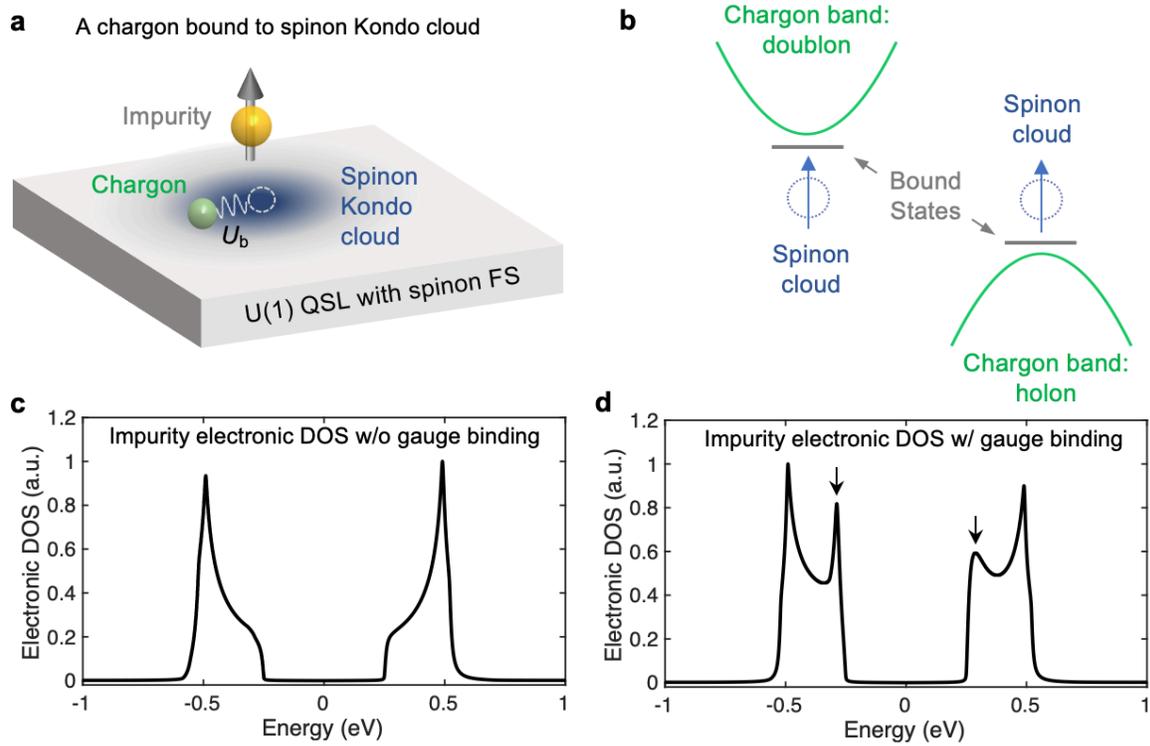

**Fig. 5. Gauge-field fluctuations induce electronic bound states. a**, Sketch shows gauge-field interaction inducing spinon-chargon bound state around a magnetic impurity. The inclusion of gauge-field fluctuations generates an effective attraction $U_b$ that binds a QSL chargon to the spinon Kondo cloud. **b**, Energy diagram illustrates the formation of spinon-chargon bound states. The attraction of chargons to the spinon Kondo cloud causes bound states to form at the Hubbard bandedge energies. **c**, The impurity electronic DOS spectrum without gauge-field fluctuations (same as Fig. 4e) ($U_b = 0$ eV). **d**, Inclusion of spinon-chargon binding induces side resonance peaks corresponding to Kondo-induced bound states in the impurity electronic DOS at the Hubbard bandedge energies (arrows) ($U_b = 0.19$ eV).